\documentclass[aip,rsi,reprint,numerical,twocolumn]{revtex4-1}
\usepackage{graphicx}
\usepackage{dcolumn}
\usepackage{bm}
\begin{document}

\title{Capturing ultrafast magnetic dynamics by time-resolved soft x-ray magnetic circular dichroism}

\author{Kou Takubo}
\email{ktakubo@issp.u-tokyo.ac.jp}
\affiliation{Institute for Solid State Physics, University of Tokyo, Kashiwa 277-8581, Japan}
\author{Kohei Yamamoto}
\affiliation{Institute for Solid State Physics, University of Tokyo, Kashiwa 277-8581, Japan}
\author{Yasuyuki Hirata}
\affiliation{Institute for Solid State Physics, University of Tokyo, Kashiwa 277-8581, Japan}
\author{Yuichi Yokoyama}
\affiliation{Institute for Solid State Physics, University of Tokyo, Kashiwa 277-8581, Japan}
\author{Yuya Kubota}
\affiliation{Institute for Solid State Physics, University of Tokyo, Kashiwa 277-8581, Japan}
\author{Shingo Yamamoto}
\affiliation{Institute for Solid State Physics, University of Tokyo, Kashiwa 277-8581, Japan}
\author{Susumu Yamamoto}
\affiliation{Institute for Solid State Physics, University of Tokyo, Kashiwa 277-8581, Japan}
\author{Iwao Matsuda}
\affiliation{Institute for Solid State Physics, University of Tokyo, Kashiwa 277-8581, Japan}
\author{Shik Shin}
\affiliation{Institute for Solid State Physics, University of Tokyo, Kashiwa 277-8581, Japan}
\author{Takeshi Seki}
\affiliation{Institute for Materials Research, Tohoku University, Sendai 980-8577, Japan}
\author{Koki Takanashi}
\affiliation{Institute for Materials Research, Tohoku University, Sendai 980-8577, Japan}
\author{Hiroki Wadati}
\affiliation{Institute for Solid State Physics, University of Tokyo, Kashiwa 277-8581, Japan}

\date{\today}

\begin{abstract}
Experiments of time-resolved x-ray magnetic circular dichroism (Tr-XMCD) and resonant x-ray scattering at a beamline BL07LSU in SPring-8 with a time-resolution of under 50 ps are presented. 
A micro-channel plate is utilized for the Tr-XMCD measurements at nearly normal incidence both in the partial electron and total fluorescence yield (PEY and TFY) modes at the $L_{2,3}$ absorption edges of the 3$d$ transition-metals in the soft x-ray region.
The ultrafast photo-induced demagnetization within 50 ps is observed on the dynamics of a magnetic material of FePt thin film, having a distinct threshold of the photon density.
The spectrum in the PEY mode is less-distorted both at the $L_{2,3}$ edges compared with that in the TFY mode and has the potential to apply the sum rule analysis for XMCD spectra in pump-probed experiments.  
\end{abstract}

\maketitle

Control of electron, magnetic, and lattice states by optical excitations in magnetically ordered materials has attracted considerable attention due to their potential applications in electronic and magnetic recording media functioning on an ultrafast time scale below nanosecond (ns, 10$^{-9}$ second, GHz range), since the observation of the ultrafast demagnetization in Ni within 1 picosecond (ps).\cite{Beaurepaire96}
The ultrafast photo-induced changes of magnetic states are non-equilibrium phenomena and several mechanisms have been proposed to understand them.\cite{Nasu,Kirilyuk10,Koopmans}
These phenomena generally involve fast structural changes near surface region \cite{Thomsen,Jal} and therefore require cooperative effects, inevitably having a threshold of the photon density.

To capture their non-equilibrium dynamics, ultrafast time-resolved experiments have been carried out using ultra-short laser pulses.\cite{Kirilyuk10}
Recently, the development of a bunched synchrotron light source,\cite{Saes03,Cavalleli05,Stamm07,Boeglin10,Holldack14} and x-ray free electron lasers\cite{Yamamoto15,Higley16,Bostedt16} have enabled investigation of the dynamic phenomena with element selectivity by tuning the photon energy of the x-ray to the absorption edges of the constituent elements.
Especially, time-resolved soft x-ray spectroscopy has many unique characteristics, such as element specificity, chemical specificity and surface sensitivity, which make them versatile for application in a wide range of scientific fields including spintronics and environmental science. 
Owing to its importance, various time-resolved soft x-ray spectroscopy studies have been carried out.\cite{Holldack10,Wietstruk11,Radu11,Eschenlohr13,Tsuyama16,Ogawa12,Yamamoto13}
The time-resolved x-ray magnetic circular dichroism (Tr-XMCD) and resonant soft x-ray scattering (Tr-RSXS) measurements for magnetic and electronic materials have been conducted in
LCLS,\cite{Higley16,Bostedt16} ALS,\cite{Stamm07} and BESSY II slicing facilities.\cite{Boeglin10,Holldack10,Radu11,Wietstruk11,Holldack14,Eschenlohr13,Tsuyama16}
The time-resolved x-ray magneto optical Kerr effect (XMOKE) measurements have also been conducted in FERMI.\cite{Yamamoto15}
Tr-XMCD measurements at the $M$ absorption below $hv<$100 eV have also been performed by using high harmonics generation from tabletop lasers in recent years.\cite{Kfir}

XMCD at the $L_{2,3}$ absorption of 3$d$ transition-metals in the soft x-ray region ($hv>$ 400 eV) is
a powerful tool to evaluate the magnetic moments at the specific sites.\cite{Chen90,Thole92}
Sum rules for XMCD at the $L_{2,3}$ edges in x-ray absorption spectroscopy (XAS) allow us to determine the orbital and spin contributions to the magnetic moments.  
The static soft XAS and XMCD in the total electron yield (TEY) mode are the most functional and simple methods, in which the photo-current induced by x-ray absorption is measured.
In the pump-probe experiments, however, the photo-current induced by the pump laser exceeds that by the probe x ray.
Thus, previous pump-probe XAS and XMCD experiments
have been performed in the transmission setting for thin films on transmissive substrates or foils like Si$_3$N$_4$, Al, and so forth.\cite{Cavalleli05,Stamm07,Boeglin10,Radu11,Eschenlohr13}
However, the good qualities of magnetic materials such as a perpendicular magnetic FePt thin film are unable to be achieved on the transparent Si$_3$N$_4$.
The x-ray reflectivity and XMOKE measurements have also been performed for non-transparent thin films or bulk samples but with rather grazed settings for their magnetized axis.\cite{Holldack14,Yamamoto15,Tsuyama16}

Here, we presents a setting for Tr-XAS and Tr-XMCD in the partial electron yield (PEY) and total fluorescence yield (TFY) modes to measure non-transmissive as-grown samples at nearly normal incidence.
The PEY mode, in which the emitted photoelectrons are measured, is rather surface sensitive but these spectra are known to usually be similar to those obtained in the TEY.\cite{Lau02}
On the other hand, the TFY mode, in which the fluorescence of an x-ray is measured, is rather bulk sensitive compared to EY.
Despite its bulk-sensitivity, the relationship between FY and the absorption coefficient is non-trivial and saturation effects become important in the TFY.\cite{Troger92,Achkar11}
The saturation effects in the TFY often change the ratio of XMCD at the $L_2$ and $L_3$ edges and preclude applying the sum rule analysis to them.

\begin{figure*}[t!]
	\includegraphics[clip]{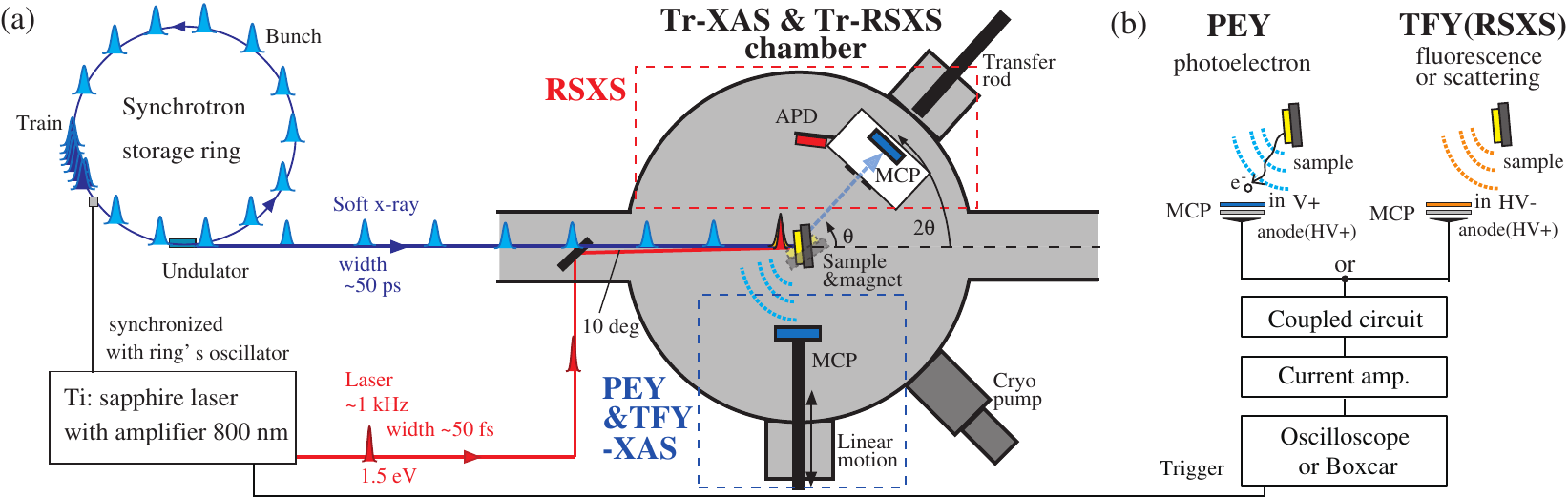}%
	\caption{(a) Overview of the setup for Tr-XAS and Tr-RSXS measurements at BL07LSU of SPring-8.
		The Ti-sapphire laser with a $\sim$1 kHz repetition rate, which is synchronized with the bunches of the synchrotron and delayed electronically, is introduced into the experimental chamber.
		The RSXS measurements are preformed at the $ 0^\circ<\theta<90^\circ$ side of the chamber (red dashed box). $\theta$ is defined as the angle between the x-ray and sample surface.
		The XAS and XMCD measurements in the PEY or TFY mode are performed at the $\theta>90^\circ$ side (blue dashed box). 
		(b) Electrical circuits for the detectors of MCP and APD. The oscilloscope or Boxcar integrator is triggered by the repetition of the laser and used for the data accumulation. 
	}
\end{figure*}

FePt thin films have drawn intense research interest owing to their potential for high density recording applications by using their magnetism.\cite{Platt02,Shima02,Seki11,Bedanta15,Iwama16}
FePt has a high uniaxial magnetic anisotropy when the L1$_0$ ordered structure is formed, 
and exhibits perpendicular magnetization in L1$_0$-FePt (001) thin films, which makes it a candidate for heat-assisted magnetic recording (HAMR),\cite{Weller00} which is an already-working technology that incorporates laser heating into the recording process.
Recently, it has been shown that proper heat-sink layers optimize the recording-time window down to 0.2 ns in HAMR of FePt.\cite{Weller15}
The challenge of achieving shorter recording-time has led to an intensive research of the ultrafast magnetization dynamics.
Very recently, Lambert \textit{et al.} have showed that circularly polarized laser pulses can induce a small helicity-dependent magnetization in FePtAgC granular film.\cite{Lambert14}
This finding suggests the possibility of all-optical switching of the magnetism in FePt having very short recording-time windows on the ps time scale.
However, the debate on the origin of this effect is on-going, and more experiments are necessary to clarify if FePt could be switched on the ultrafast time scale.\cite{Nieves16}

In this letter, the results of Tr-XMCD experiments for as-grown FePt films with non-transmissive substrates are reported.
By virtue of the pump-probe technique, a photo-induced ultrafast demagnetization at the Fe sites of the FePt thin film within the experimental time-resolution of 50 ps for Tr-XMCD at SPring-8.

Figure 1 shows an overview of the experimental setup for Tr-XAS, Tr-XMCD,
and Tr-RSXS measurements in the soft x-ray region at BL07LSU of SPring-8.\cite{Yamamoto14}
Tr-RSXS is performed at the $\theta < 90^\circ$ side of the experimental chamber.
$\theta$ is the angle between the x-ray and sample-surface as the standard definition of x-ray diffraction.  
The scattering is detected by the micro-channel plate (MCP) or avalanche photodiode (APD) installed on the 2$\theta$ motion of the diffractometer.\cite{future}
On the other hand, XAS and XMCD in the PEY or TFY modes are measured at the $\theta > 90^\circ$ side of the chamber.
Emitted photoelectrons or x-ray fluorescence are caught by another MCP topped on the linear motion, which could be positioned as close as $\sim$2 cm from the sample surface.
The femtosecond Ti:sapphire laser with a wavelength of $\sim$ 800 nm housed at the laser station of BL07LSU\cite{Ogawa12,Yamamoto13,Yamamoto14} is introduced into the XAS and RSXS chamber.
The laser irradiates samples 10$^\circ$ below the x-ray and photo-induced dynamics of the electronic and structural evolutions are examined by means of a pump-probed technique.
The laser pulses with $\sim$ 1 kHz repetition rate are synchronized with selected bunches of the synchrotron and delayed electronically.
The pulse width of the Ti:sapphire laser is $\sim$ 50 fs.
On the other hand, the single bunch width of the x-rays in the H-mode and F-mode at SPring-8\cite{SPring8} is $\sim$ 50 ps, which limits the experimental time-resolution.
Figure 2 shows an example of the time-profile of the signals in the F-mode.
The signals of the MCP or APD are amplified and gated on the oscilloscope or Boxcar integrator which is triggered by the laser pulses [Fig.1 (b)].
 
A chevron-type (dual-plate) MCP \cite{Wiza79} is leveraged for XAS and XMCD measurements in the PEY and TFY modes.
Each surface channel of the MCP detects both charged particles (photoelectrons) and energetic
photons (x-ray) and amplifies them as the electric currents.
When a positive voltage is applied to MCP-in terminal of the surface plate,
both the photoelectrons and x-ray fluorescence are detected, as illustrated in Fig. 1(b).
However, since the current induced by the photoelectron is much larger than that by x-ray,
the spectrum can be regarded to arise from the PEY.
Alternatively, when a negative high voltage (HV) is applied to the MCP-in,
the photoelectrons are bounded and only the fluorescence reaches to MCP.
In this case, the TFY spectrum is obtained.

\begin{figure}[]
	\includegraphics[clip]{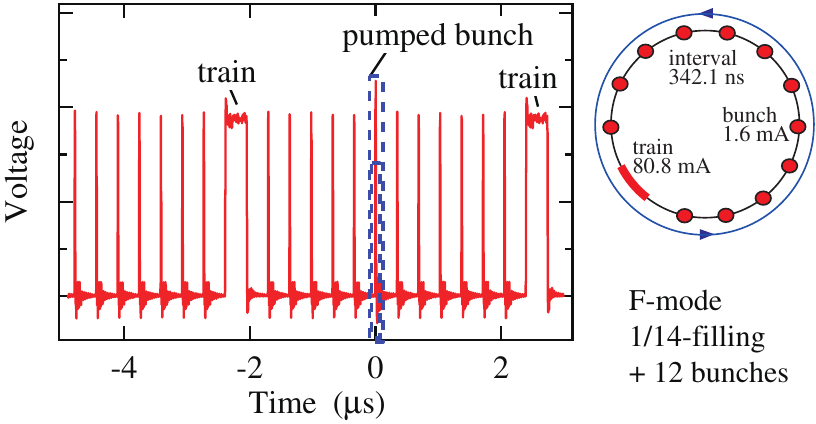}%
	\caption{
		An example of signal output observed on the oscilloscope during the time-resolved measurements
		in the F-mode at SPring-8.\cite{SPring8} The example corresponds to the data at $t$=30 ps taken with the $\mu^-$ polarization, shown in Fig.4 (a). The dashed box denotes the gated region.
		Right panel shows the bunch configuration of the F-mode.
	}
\end{figure}

A FePt (001) thin film with the thickness of 50 nm was epitaxially grown at 500$^\circ$C on an MgO (100) single crystal substrate in an ultrahigh vacuum magnetron sputtering system.\cite{Seki11}
The high temperature deposition process promoted the L1$_0$ ordering of FePt, resulting in the perpendicular magnetization.
Right and left circularly polarized
x-rays ($\mu^+$ and $\mu^-$) for the XMCD measurement were obtained by combinations of horizontal and vertical figure-8 undulators in BL07LSU.\cite{Yamamoto14}
The x-ray beams were focused using a mirror and its size was 100 $\mu$m $\times$ 200 $\mu$m on the sample.
The magnetic field was applied by a 260mT permanent magnet placed behind the substrate.
XMCD was measured at 10 degrees off normal incidence ($\theta$ = 100$^\circ$).
The laser was linearly polarized and focused to a size $\sim$ 1mm$^2$ on the sample.

Figure 3 shows the static XMCD spectra at the Fe $L_{2,3}$ edges for the FePt thin film in the TEY, PEY, and TFY modes.
Intense XMCD was observed both at the $L_{2,3}$ edges, $\sim$-40$\%$ at 707 eV and $\sim$6$\%$ at 720.2eV for the original $\mu^-$ XAS in TEY and PEY.
A spectral difference between the TEY and PEY mode is barely observable.
On the other hand, XMCD at the $L_2$ edge in the TFY mode is scarcely observed owing to the distortion caused by the saturation effect, although TFY has a bulk-sensitivity.  
Using the sum rules, the magnetic moments at the Fe site are estimated to be $m_{spin}$ = 2.63$\mu_B$ and $m_{orbital}$ = 0.10$\mu_B$ for TEY, and $m_{spin}$ = 2.74$\mu_B$ and $m_{orbital}$ = 0.15$\mu_B$ for PEY, respectively.
Here, the number of the 3$d$ electrons was assumed to be $n_{3d}$=6.6.
These values are basically consistent with previous studies,\cite{XMCD1,XMCD2,XMCD6} while it has already been argued that there were some systematic discrepancies arising from the errors in the background subtraction procedures.\cite{XMCD6} 
On the other hand, the magnetic moments are estimated to be $m_{spin}$ = 1.53$\mu_B$ and $m_{orbital}$ = 0.08$\mu_B$ for the TFY spectra, which exhibited a large discrepancy arising from their strong distortion.

\begin{figure}[]
\includegraphics[clip]{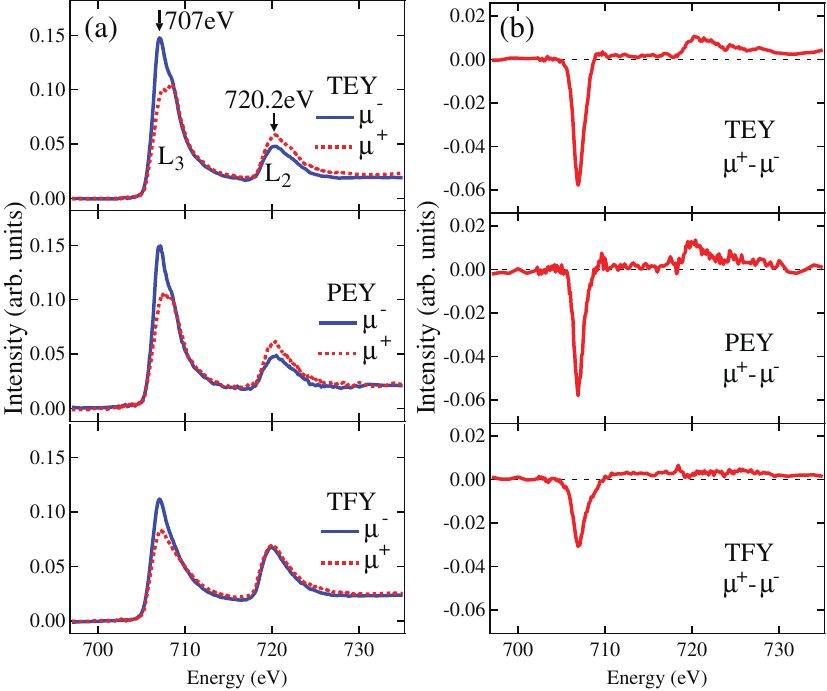}%
\caption{
Static XMCD spectra of the FePt thin film taken in the TEY (top), PEY (middle), and TFY (bottom) modes at room temperature.
(a) XAS spectra taken at the $\mu^+$ and $\mu^-$ polarization.
The spectra are normalized by the area between 700 and 740 eV.
The arrows denote the energies used in the time-resolved measurements for the $L_{2,3}$ edges.   
(b) XMCD intensities.
}
\end{figure}

The photo-induced dynamics of the FePt thin film were examined in the less-distorted PEY mode.
The time-evolutions of the FePt thin film with 16 mJ/cm$^2$ laser irradiation are given in Fig. 4 (a) at $h\nu$=707.0 eV ($L_3$ edge) and (b) at 720.2 eV ($L_2$ edge), respectively.
As can be seen, almost similar time-evolutions are observed for the Fe $L_{2,3}$ edges.
XMCD, namely the difference of the intensities at $\mu^+$ and $\mu^-$, decreases immediately after the pumped pulse irradiation both at the Fe $L_{2,3}$ edges.
XMCD exhibits a reduction in its intensity by $\sim$90$\%$ of the original value at both the $L_{2,3}$ edges $\sim$30 ps after the pump pulse.
Then the XMCD exhibits a slow recovery of the magnetization.
The time-evolutions are fitted by the following function
\begin{equation}
I(t) = I_1 \exp(-t/\tau_{fast}) + I_2 [1-\exp(-t/\tau_{relax})],
\end{equation}
convoluted with the Gaussian response function of the time-resolution ($\tau_{Gauss}$=50 ps). 
The 4 time-evolutions can be fitted with similar time constants of $\tau_{fast}=25\pm10$ ps and $\tau_{relax}=150\pm10$ ps [solid lines in Fig. 4(a) and (b)].
On analysis, the photo-induced demagnetization occurs within the experimental time resolution $<$ 50 ps after the pump pulse, and is relaxed with a time constant of $\sim$150 ps.
It should be noted that the time constant of $\sim$ 150 ps for the demagnetization recovery in the FePt thin film is much slower than that observed in the elementary metals Fe, Ni, and so forth of $\sim$10 ps, \cite{Kirilyuk10} and similar to the slow recovery of several hundred ps time scale observed for the intermetallic GdFeCo\cite{Kirilyuk10,Radu11} and Co/Pt multi layers,\cite{Kanzantseva08} which also show magnetization reversals induced by the laser pulses.
In addition, the magnetization is not recovered to the original value even at $t$=1500 ps, which is completed before the next bunch arrives after $\sim$342 ns.
The time constant of $\sim$ 150 ps will correspond to the time for spin and electron temperature transferring into the lattice, as discussed below.

\begin{figure}[]
	\includegraphics[clip]{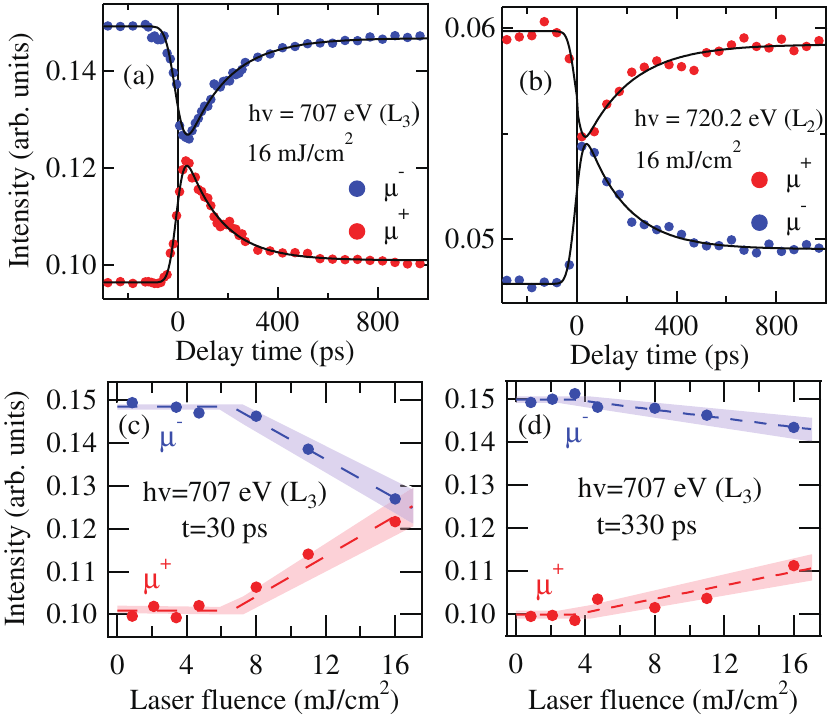}%
	\caption{
		Time-evolutions of XMCD taken in the PEY mode at (a) $L_{3}$ ($h\nu$=707.0 eV) and (b) $L_2$ ($h\nu$=720.2 eV) edges for the FePt thin film at room temperature.
		The solid lines denote the fitting results (see text).
		(c)(d) Laser-fluence dependence of Tr-XMCD intensities ($\mu^+$ and $\mu^-$) at (c) $t$=30 ps and (d) $t$=330 ps on the $L_{3}$ edge ($h\nu$=707.0 eV).
		The dashed lines denote results of the linear fittings having a threshold. 
		Shades indicate the error of the fittings.
	}
	
\end{figure}

Figure 4 (c)(d) gives the Tr-XMCD intensities ($\mu^+$ and $\mu^-$) at (c) $t$=30 ps and (d) $t$=330 ps as a function of the laser fluence.
Threshold-like behaviour is observed at 6.5 mJ/cm$^2$ for $t$=30 ps, which is estimated by the linear fittings for the fluence dependences.
The threshold seems to decrease as the time $t$ increases (See supplementary for more information).
These results indicate that the photo-induced effect will not be a simple thermal effect.\cite{Nasu,Tsuyama16} 
The threshold-like behaviour, namely the abrupt increase of the change ratio at around a few tens ps scale with high fluence seems to be a consequence of the critical magnetization fluctuation near the Curie temperature,
which was suggested in previous Tr-MOKE studies.\cite{Mendil,Kimling}
The demagnetization dynamics for general metallic magnets have been described by a so-called three-temperature model\cite{Kirilyuk10,Kimling,Koopmans} in the Tr-MOKE studies, while Tr-MOKE detects macroscopic magnetizations and does not have an element selectivity.
Sub-ps demagnetization and following ps remagnetization are observed as the first step of the dynamics, that may be hindered by the present temporal resolution.
Then sub-ps de- or remagnetization is followed by a slower step (extend to $\sim$several hundred ps) as the lattice temperature increases, that is clearly seen in the present dynamics. 
One may consider that all of demagnetization processes can be described by temperature dependences of the electron and spin specific heats and does not relate to a cooperative effect including the 'lattice'.
However, the photo irradiation also causes a sudden volume expansion near the surface just after the phonon temperature becomes to raise, which is usually lasting for several hundred ps.\cite{Thomsen}
In addition, the slow relaxation of FePt with time constant of $\sim$ 150 ps indicates that the 'lattice' temperature is also important especially on this time scale (See supplementary again).
Furthermore, a recent time-resolved x-ray resonant magnetic reflectivity measurement revealed that demagnetization dynamics in a Ni film involves a structural change near the surface region.\cite{Jal}
Therefore, it will be reasonable that the photo-induced demagnetization of FePt involves a kind of lattice cooperative effect and is not a simple thermal effect.
Further time-resolved structural study including x-ray diffraction and reflectivity measurements will be very suggestive in order to clarify this point, which can be performed in same chamber \textit{in situ}.

Here, we report the advantages of the present Tr-XAS and XMCD measurements, which are (1) the precise sum rule analysis in the PEY mode at nearly normal incidence, and (2) the depth-dependent measurements utilizing the difference of probing depths between PEY and TFY.
When some negative retarding voltages are applied to the MCP-in terminal in the PEY mode, the surface sensitivity can be tuned between about 1 nm and 10 nm.
On the other hand, the TFY mode can detect the bulk region of $\sim$100 nm.
The depth-resolved dynamical measurements will be archived by using this technique.
The difference of photo-induced dynamics between surface and bulk region on several materials including surface and/or interface magnetisms will be clarified.
The less-distorted PEY method will provide an opportunity for applying a sum rule analysis to XMCD spectra in the ultrafast pump-probed spectroscopy for various magnetic materials,
which would be very useful to explore the all-optical switching in spintronics devices such as FePt, Co/Pt multi-layers and so forth.  
On the other hand, the bulk sensitive TFY method will allow us to measure the dynamics of bulk samples showing various quantum phenomena including high-$T_c$ superconductors, heavy fermion metals and so forth.
In conclusion, we have studied the photo-induced magnetic dynamics of the FePt thin films at the Fe site using the Tr-XMCD technique at nearly normal incidence. 
The ultrafast photo-induced demagnetization within 50 ps and its slow relaxation taking a few hundred ps
are clarified with a distinct threshold-like behaviour.
MCP has been used for Tr-XMCD both in the electron and fluorescence yield modes at the $L_{2,3}$ edges of the 3$d$ transition-metals in the soft-x-ray region.
The spectrum in the PEY mode is less distorted and basically similar to that obtained in the TEY mode.
No difference of the change ratio is observed between the Fe $L_2$ and $L_3$ edges in the time-resolved measurement, taking advantage of the PEY method.

See supplementary material for additional data of laser fluence dependence at various pump-probe delays.

We acknowledge kind support by Dr. Y. Tanaka, Dr. T. Ohkochi and Prof. T. Kinoshita.
Tr-XMCD measurements were performed with the approval of Synchrotron Radiation Research Organization, the University of Tokyo (No. 2016A7504, 2016A7403, 2016B7403, 2016B7518). 
This research was supported by the Japan Society the Promotion of Science (JSPS) of Grant-in-Aid for Young Scientists (B) (Nos. 16K20997 and 16K17722) and for Scientific
Research (C) (No. 26400328).
This work was also supported by the Ministry of Education, Culture, Sports, Science and Technology of Japan (X-ray Free Electron Laser Priority Strategy Program).

\clearpage

\onecolumngrid
\appendix

\renewcommand{\bibname}{References}
\renewcommand{\thefigure}{S\arabic{figure}}

\vspace{6mm}

\begin{center}
	\huge{Supplemental Material}
	
	\vspace{0.6cm}
	
	\Large{\textbf{Capturing ultrafast magnetic dynamics by time-resolved soft x-ray magnetic circular dichroism}}
	
	\vspace{0.6cm}
	
	\large{Kou Takubo,$^1$ Kohei Yamamoto,$^1$ Yasuyuki Hirata,$^1$ Yuichi Yokoyama,$^1$ Yuya Kubota,$^1$\\ Shingo Yamamoto,$^1$, Susumu Yamamoto,$^1$ Iwao Matsuda,$^1$ Shik Shin,$^1$ Takeshi Seki,$^2$\\ Koki Takanashi,$^2$ and Hiroki Wadati$^1$}
	
	\vspace{4mm}

	\normalsize{$^{1}$Institute for Solid State Physics, University of Tokyo, Kashiwa 277-8581, Japan}\\
	\normalsize{$^{2}$Institute for Materials Research, Tohoku University, Sendai 980-8577, Japan}\\
	
	\vspace{5mm}
	
\end{center}


\begin{figure}[h!]
	\includegraphics[width=1\linewidth]{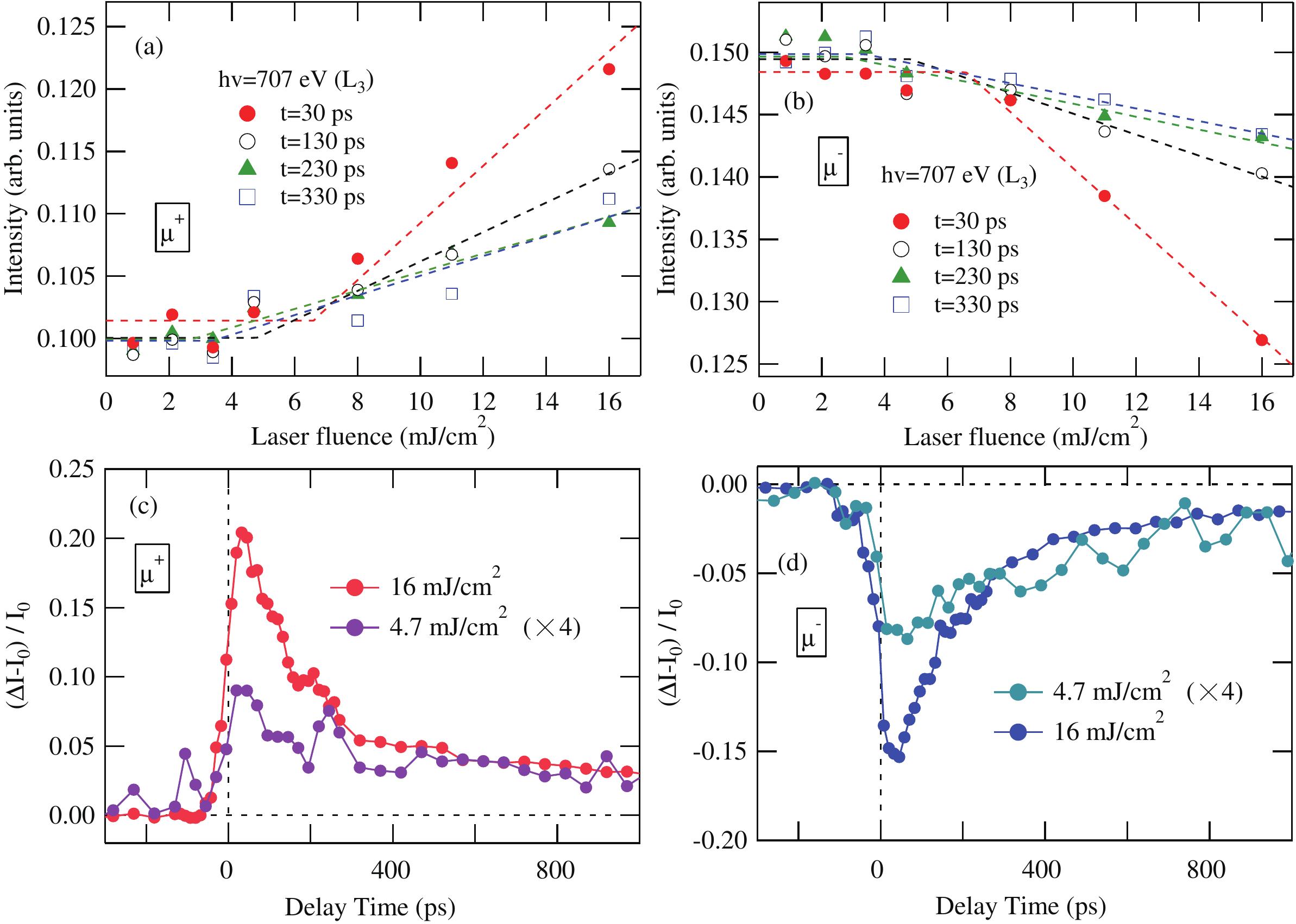}%
	\caption{
		(a)(b) Laser-fluence dependence of Tr-XMCD intensities (a) $\mu^-$ and (b) $\mu^+$ at various time delays ($t$= 30, 130, 230, and 330 ps) on the $L_{3}$ edge ($h\nu$=707.0 eV) of the FePt thin film.				
		Dashed lines indicate the results of linear fittings with thresholds.
		(c)(d) Time-profiles of the Tr-XMCD intensities for (c) $\mu^+$ and (d) $\mu^-$ at the laser fluence of 4.7 mJ/cm$^2$ and 16.0 mJ/cm$^2$.
	}
\end{figure}

Figures S1 (a)(b) give the laser-fluence dependence of Tr-XMCD intensities (a) $\mu^+$ and (b) $\mu^-$ at various time delays.
Their thresholds are estimated by linear fittings (dashed lines) for the various delays; 6.5$\pm$0.7 mJ/cm$^2$ for $t$=30 ps, 4.8$\pm$0.8 mJ/cm$^2$ for $t$=130 ps, 2.7$\pm$1.0 mJ/cm$^2$ for $t$=230 ps,
and 3.4$\pm$0.9 mJ/cm$^2$ for $t$=330 ps, respectively, which are denoted by the crossing point of the dashed lines.
The data for $t$=30 ps and $t$=330 ps are also shown in Fig. 4 (c)(d) of the main text.
The threshold exhibits a largest value of 6.5 mJ/cm$^2$ at $t$=30 ps, and decreases as the time $t$ increases.
And then it seems to be constant $\sim$3 mJ/cm$^2$ after $t$$\sim$230 ps.
$t\sim$230 ps seems to correspond to the time when the phonon temperature finishes to increase, as discussed later.
The threshold-like behaviour, namely the abrupt increase of the change ratio at around a few tens ps scale with high fluences seems to be a consequence of the critical magnetization fluctuation near the Curie temperature, which was suggested in Ref. [S1] and [S2].
On the other hand, the monotonic increases shown in Fig. S1(a)(b) in wide ranges of the fluence between $\sim$3 and 16 mJ/cm$^2$ will be difficult to be described by an uniformed temperature model.

Figures S1 (c)(d) give the time-profile of Tr-XMCD for (c) $\mu^+$ and (d) $\mu^-$ at the laser fluence of 4.7 mJ/cm$^2$ and 16.0 mJ/cm$^2$ on the $L_{3}$ edge ($h\nu$=707.0 eV).
The dynamics at lower fluence of 4.7 mJ/cm$^2$ and higher fluence of 16.0 mJ/cm$^2$ seem to exhibit different profiles.
The change ratio of $\sim$4.7 mJ/cm$^2$ at $t$$\sim$30 ps in which it take the maximum is relatively small compared to that at around $t$$\sim$200 ps.
Both of the time-evolutions are not recovered to the original value even at $t$=1000 ps.
Therefore, the relaxation time constant of $\sim$ 150 ps estimated in the main text will correspond to the time of the spin and electron temperature transferring into the lattice, namely the speed that the lattice temperature increases.
Then, after $\sim$ 150 ps, the ‘three temperatures’ of the electron, spin, and lattice will reach a same temperature, and finally are relaxed slowly taking a few hundreds ns.
These slow increases of the lattice temperature will relate to the sudden structural expansion near the surface [S3,S4], as discussed in main text.
The photo irradiation can cause a fast volume expansion near the surface region just after the phonon temperature becomes to raise. And it causes so-called 'shockwaves' which were observed for various materials without phase transitions and lasting several hundreds ps (See Ref. [S3]).

\section*{References}

[S1] J. Mendil, P. Nieves, O. Chubykalo-Fesenko, J. Walowski, T. Santos, S. Pisana, and M. M\"{u}nzenberga, Sci. 

\hspace{7mm}Rep. {\bf4}, 3980 (2014).

[S2] J. Kimling, J. Kimling, R. B. Wilson, B. Hebler, M. Albrecht, and D. G. Cahill, Phys. Rev. B {\bf 90}, 224408

\hspace{7mm}(2014).

[S3] C. Thomsen, H. T. Grahn, H. J. Maris, and J. Tauc, Phys. Rev. B {\bf 34}, 4129 (1986).

[S4] E. Jal, V. L\'{o}pez-Flores, N. Pontius, C. Sch\"{u}{\ss}ler-Langeheine, T. Fert\'{e}, N. Bergeard, C. Boeglin, B. Vodungbo,

\hspace{7mm}J. L\"{u}ning, and N. Jaouen, arXiv:1701.01375v1.


\begin{references}
\bibitem{future}
Results of the Tr-RSXS measurements will be reported elsewhere.
\end{references}
\end{document}